\documentclass[11pt]{article}

\usepackage{a4wide}
\usepackage{amsmath,amssymb,amsthm,amscd}
\usepackage[pdftex]{graphicx}

\newcommand{\bt}{\pmb\theta}

\newcommand{\vc}[1]{{\bf #1 }}
\newcommand{\const}{\operatorname{const}}
\newcommand{\Var}{\operatorname{Var}}
\newcommand{\Cov}{\operatorname{Cov}}
\newcommand{\tr}{\operatorname{tr}}

\def\identy{{\mathsurround0pt\mathchoice{\textidenty}{\textidenty}{\scptidenty}{\scptidenty}}}
\def\scptidenty{\setbox0\hbox{$\scriptstyle1$}\bothidenty}
\def\textidenty{\setbox0\hbox{$1$}\bothidenty}
\def\bothidenty{\rlap{\hbox to.97\wd0{\hss\vrule height.06\ht0 width.82\wd0}}
 \copy0\rlap{\kern-.36\wd0\vrule height1.05\ht0 width.05\ht0}\kern.14\wd0}

\begin{document}

\title{A Simulated Annealing Approach to Bayesian Inference}

\author{Carlo Albert\footnote{Eawag, Swiss Federal Institute of Aquatic Science and Technology, 8600 D\"ubendorf, Switzerland.}}

\maketitle

\abstract{
      A generic algorithm for the extraction of probabilistic (Bayesian) information about model parameters from data is presented.
    The algorithm propagates an ensemble of particles in the product space of model parameters and outputs. Each particle update consists of a random jump in parameter space followed by a simulation of a model output and a Metropolis acceptance/rejection step based on a comparison of the simulated output to the data.
    The distance of a particle to the data is interpreted as an energy and the algorithm is reducing the associated temperature of the ensemble such that entropy production is minimized.
    If this simulated annealing is not too fast compared to the mixing speed in parameter space, the parameter marginal of the ensemble approaches the Bayesian posterior distribution.
    Annealing is adaptive and depends on certain extensive thermodynamic quantities that can easily be measured throughout run-time.
    In the general case, we propose annealing with a constant entropy production rate, which is optimal as long as annealing is not too fast.
    For the practically relevant special case of no prior knowledge, we derive an optimal fast annealing schedule with a non-constant entropy production rate.
    The algorithm does not require the calculation of the density of the model likelihood, which makes it interesting for Bayesian parameter inference with stochastic models, whose likelihood functions are typically very high dimensional integrals.
}

\section{Introduction}

While the amount of data measured in various complex systems keeps growing we are in need of algorithms that allow us to extract useful information from this data.
If we manage to set up a model, of which the data appears to be a realization, this information can conveniently be expressed as a distribution of model parameters that is "compatible" with the data.
If the model is to be used to make reliable predictions, it is important for it to faithfully reproduce the relevant statistics of the data (e.g. frequency of extreme events), which naturally leads us to stochastic models with non-trivial output distributions.

{\em Bayesian statistics} is a framework in which a stochastic model is expressed through a conditional probability distribution ({\em likelihood function}), $L(\vc x|\bt)$, for observable outputs $\vc x$, given model parameters $\bt$, and knowledge (or belief) about model parameters is expressed through a probability distribution as well.
Assume we have prior knowledge in the form of a probability distribution, $f_{pri}(\bt)$, and measured data, $\vc y$, which is believed to be a realization of the model.
The posterior knowledge, combining prior knowledge with the one acquired from data, is calculated by means of eq.
\begin{equation}\label{posterior}
f_{post}(\bt|\vc y)=\frac{L(\vc y|\bt)f_{pri}(\bt)}{\int L(\vc y|\bt')f_{pri}(\bt')d\bt'}\,.
\end{equation}
This so-called {\em Bayes theorem} is consistent with sequential learning.
The posterior becomes the new prior when new data becomes available, and the posterior $f_{post}(\bt|\vc y_1,\dots,\vc y_n)$ does not depend on the order in which the data points $\vc y_1$ through $\vc y_n$ are learned from.

Since (\ref{posterior}) is usually intractable, analytically, Bayesian inference algorithms usually aim at generating, by means of Monte Carlo methods, a sufficiently large parameter sample from it.
This sample expresses our posterior knowledge about model parameters.
The maximum of the posterior constitutes our best guess about model parameters. It is a compromise between our prior knowledge and the best model fit of the data.
But we are really interested in the whole posterior distribution, as it expresses the {\em uncertainty about model parameters}.
Predictive uncertainty can conveniently be estimated through sampling one (or several) model outputs, for each posterior parameter sample point, from an appropriate likelihood function.
For a recent review on Bayesian inference in physics see \cite{vonToussaint2011bayesian}.

In cases where the likelihood function can be evaluated in reasonable time, the {\em Metropolis algorithm} \cite{metropolis_1953_MRT2} with its various ramifications is an efficient way of generating a posterior parameter sample.
Usually, first, a global optimizer is employed to determine the maximum of the posterior, which serves as the starting point for the Metropolis algorithm.
In practice, evaluating $L(\vc y|\bt)$, for tens of thousands of parameter combinations, which is typically required for both the optimization and the Metropolis algorithm, is often not feasible.
If evaluating $L(\vc y|\bt)$, for given $\vc y$ and $\bt$, is slow, we have to distinguish two cases, depending on whether {\em simulating} a model output, for given $\bt$, is slow or fast.

As a typical example for the first case consider a slow {\em deterministic} simulation model, $\vc g(\bt)$, together with a multivariate normal error with mean zero that is added to this deterministic model output.
In practice, such normal errors are often used and supposed to lump the effects of input, model structural and measurement uncertainties \cite{reichert_2012_bias}.
Simulating a model output and evaluating the model likelihood for the data, for a given $\bt$, are then more or less of the same cost. Both require the calculation of $\vc g(\bt)$. In addition, to simulate a model output, we need to draw from a normal distribution, whereas to evaluate the model likelihood we need to evaluate the associated normal density. The cost of these last steps is typically negligible compared to the calculation of $\vc g(\bt)$.
If one wants to make a Bayesian inference with such models, one needs to think of ways how to speed up the calculation of $\vc g(\bt)$ \cite{ohagan_2006_EmulatorTutorial, reichert_2011_dynemulator, albert_2012_emulator}.

As a typical example for the second case, consider a dynamical model that is defined through a stochastic differential equation (SDE), and whose output is a time-series.
Simulating an output with such a model is often very fast; it simply requires the generation of a realization of the stochastic process defined by the SDE.
On the other hand, evaluating the likelihood, for a given measured time-series, is typically very expensive. It requires us to numerically solve a path-integral over all possible realizations of the SDE that are compatible with the measured time-series.
It is this second kind of problems this paper is concerned with: How to do a Bayesian inference with models that allow for a quick realization of an output, but whose likelihood densities are expensive to evaluate.
Indeed, a faithful account of uncertainty in models of complex systems typically does require us to use stochastic models whose likelihood functions are very high dimensional intractable integrals.

Therefore, in recent years, there has been a great interest in so-called {\em Approximate Bayes Computations} (ABC), especially within the statistics community (see \cite{marin_2012_ABC} for a recent review).
These algorithms are based on {\em simulating outputs} from the model likelihood and comparing them with the data rather than evaluating the likelihood function, for the data.
Comparison with the data requires introducing a metric on the space of outputs as well as a {\em tolerance} within which a simulated output is deemed compatible with the data.
For a posterior that is very different from the prior it is beneficial to start with a relatively large tolerance and decrease it during the course of the algorithm.
Thereby, an ensemble of particles is iteratively updated so as to converge to the exact posterior along a series of approximations, associated with a decreasing series of tolerances.
For continuous outputs the tolerance will only asymptotically approach zero and, even for moderately high output dimensions, we will have to live with a finite tolerance that justifies the word {\em approximate} in ABC.
If the output dimension is large (such as a time-series) we need to use appropriate {\em summary statistics}. If model parameters are few, it is often possible to find few summary statistics that contain almost the same amount of information about the parameters as the original output \cite{Fearnhead_2012_ABC}.
The ABC framework is generic in the sense that it only requires us to simulate outputs, for given parameter sets. No other information about the model is required.
We do not even need access to the model equations.

Recently, a framework of ABC algorithms \cite{Albert_2013_ABC} has been introduced that is inspired by adaptive {\em simulated annealing} algorithms that were developed for optimization tasks \cite{nulton1988statistical, salamon1988simulated, andresen_1988_lumpedThdynPropSA, ruppeiner_1991_EnsembleSA, andresen_1994_CTS}.
These so-called SABC algorithms (merging SA, for simulated annealing, with ABC), attempt to move an ensemble of particles as closely as possible to the exact posterior with as few computer updates as possible.
Each fundamental update step consists of a random jump in parameter space followed by a simulation of an output and a Metropolis acceptance/rejection step that depends on a tolerance that is continuoulsy decreased during run-time.
These update steps are associated with a flow of entropy from the system (the ensemble of particles in the product space of parameters and ouputs) to the environment.
Part of this flow is due to the decrease of entropy in the system when it transforms from the prior to the posterior state and constitutes the well-invested part of computation.
Since the process happens in finite time, inevitably, additional entropy is produced. This {\em entropy production} is used as a measure of the wasted computation and minimized, as previously suggested for adaptive simulated annealing \cite{salamon1988simulated, nulton1988statistical}.
The algorithm adapts the tolerance (temperature) according to certain measurable extensive thermodynamic quantities of the system, in the easiest case simply the total distance of the particles to the target (measured data). We assume that the system, at any time, is approximately described by a Gibbs state w.r.t. these quantities.
Under this so-called {\em endoreversibility assumption} \cite{rubin_1979_endoreversibility} the entropy production is determined by the flux of the extensive thermodynamic quantities times the corresponding thermodynamic forces.
Minimization of this quantity can be carried out analytically \cite{spirkl_1995_optFiniteTimeEndorevProc} and leaves us with a family of algorithms that is basically parameterized by two parameters, governing the annealing speed and the equilibration (mixing) speed, respectively.

The convergence of adaptive SABC algorithms hinges on the endoreversibility assumption being satisfied.
If it is not, additional entropy is produced within the system. This entropy can even remain in the system indefinitely and lead to a biased convergence.
Traditional ABC algorithms \cite{marin_2012_ABC} enforce convergence through {\em importance sampling}, but the ensuing loss of effective sample size reduces their efficiency.
Furthermore, they lack an information-theoretic criterion for the speed with which the tolerance is decreased.
In \cite{Albert_2013_ABC} the endoreversibility assumption is assumed to be true if mixing in parameter space is fast enough compared to annealing.
This means that the jump width in parameter space has to be chosen sufficiently large, depending on the annealing speed.
Since, in practice, we are interested in fast annealing schedules, we further elaborate on this assumption in this paper.
We first note that, even for infinitely fast mixing, the system does not follow a sequence of Gibbs states, i.e., there is always additional internal entropy production, which is not accounted for in our optimization.
This might sound counter-intuitive, at first, but is simply due to the fact that our Metropolis particle update rule does not simulate a real gas or fluid.
However, for the practically relevant case of negligible prior information and fast annealing, we show that this additional entropy production is small compared to the entropy production caused by heat flow, as long as mixing in parameter space is fast enough.
Indeed, in this case, the system approximately follows a sequence of Gibbs states, in the limit of infinitely fast annealing and infinitely fast mixing.

In order to be self-sufficient, this paper repeats the main ideas of \cite{Albert_2013_ABC}, but with a notation that complies with physics standards.
For technical details as well as various examples that show the competitive performance of SABC algorithms the reader is referred to the original publication \cite{Albert_2013_ABC}.

\section{The SABC framework}

We define our {\em system} as a time-dependent probability distribution, $\mu(\bt,\vc x,t)$, represented by a moving ensemble of particles, $E=\{\bt_i,\vc x_i\}_{i=1}^N$, in the product space of model parameters and associated model outputs.
The {\em energy}, $u(\vc x)$, of a particle measures the distance of its output component to the data $\vc y$, w.r.t. some metric on the output space.
To simplify notation, we omit the $\vc y$-dependence of $u$.
The basic idea behind the SABC framework is to generate, through repeated model simulations, a canonical ensemble $E$ described by the distribution
\begin{equation}\label{pi}
  \pi_{T^e}(\bt,\vc x)=\frac{1}{Z(T^e)}L(\vc x|\bt)f_{pri}(\bt)e^{-u(\vc x)/T^e}\,,
\end{equation}
for a sufficiently low {\em temperature} $T^e$.
Then, the $\bt$-marginal of (\ref{pi}) will be a good approximation of the posterior (\ref{posterior}), i.e.,
\begin{equation}
  f_{post}(\bt|\vc y)
  \approx
  \int \pi_{T^e}(\bt,\vc x)d\vc x\,.
\end{equation}
If the output space is $\mathbb R^n$, the energy $u(\vc x)$ might read as
\begin{equation}\label{alpha}
    u(\vc x)=
  \rho(\vc x,\vc y)
  =
  \frac{1}{\alpha}
  \sum_{i=1}^n|x_i-y_i|^\alpha\,,
\end{equation}
for some constant $\alpha > 0$.

It is easy to write down rules for particle updates that do not require us to calculate the density of the likelihood, and which have (\ref{pi}) as equilibrium distribution (see eqs. (\ref{transition}) or (\ref{newtransition}) below).
For posteriors that are very different from the prior, it seems reasonable, in the spirit of {\em simulated annealing}, to make the temperature time-dependent, $T^e=T^e(t)$, and reduce it during the course of the algorithm.
The crucial question is then how fast we should reduce this temperature in order to have a fast convergence to the correct result.
In \cite{Albert_2013_ABC} a convergence proof has been given, for the case when annealing obeys $T^e(t) \geq\const t^{-\alpha/n}$ and the metric (\ref{alpha}) is used.
Compare this result with the logarithmic annealing \cite{granville1994simulated} that is required, for {\em optimizers} that are based on simulated annealing to converge.
However, these results are of little practical value.
What we are interested in are annealing schedules that take us as close as possible to the correct result after a finite number of particle updates.

In practice, we often lack prior knowledge about model parameters, and the posterior is mainly data driven. We derive an efficient SABC algorithm, for this special case, in the next subsection and come back to the general case at the end of this section.

\subsection{The case of negligible prior knowledge}\label{SectUninformative}

Consider the case of a prior that varies much less than the likelihood function and set
\begin{equation}
  f_{pri}(\bt)\equiv \const\,.
\end{equation}

A simple {\em transition rate}, $q_{T^e}((\bt,\vc x)\rightarrow (\bt',\vc x'))$, that has (\ref{pi}) as {\em equilibrium distribution}, i.e, satisfies the {\em detailed balance} condition
\begin{equation}\label{detbal}
  \pi_{T^e}(\bt,\vc x)q_{T^e}((\bt,\vc x)\rightarrow (\bt',\vc x')) =
  \pi_{T^e}(\bt',\vc x')q_{T^e}((\bt',\vc x')\rightarrow (\bt,\vc x))
\end{equation}
is given by eq.
\begin{equation}\label{transition}
  q_{T^e}((\bt,\vc x)\rightarrow (\bt',\vc x'))
  =
  k(\bt,\bt')L(\vc x'|\bt')\min\left(
    1,
    e^{-(u(\vc x')-u(\vc x))/T^e}
  \right)\,,
\end{equation}
where $k(\bt,\bt')$ denotes some symmetric transition rate in parameter space.
According to (\ref{transition}), a particle update consists in a random jump in parameter space, followed by the simulation of an output from the model and a {\em Metropolis acceptance/rejection step} that depends solely on the change of the particle's distance to the data.
We allow the {\em environmental temperature} to be time-dependent, $T^e=T^e(t)$.
For a sufficiently large number of particles, the dynamics of the ensemble is then approximately described by the {\em master equation}
\begin{equation}
\label{time-evol-mu}
\frac{\partial \mu(\vc z,t)}{\partial t}
=
\int(
        \mu(\vc z',t) q_{T^e(t)}(\vc z'\rightarrow \vc z)  - \mu(\vc z,t)
        q_{T^e(t)}(\vc z\rightarrow \vc z')
    )
    d\vc z'\, ,
\end{equation}
where we have used the shorthand $\vc z=(\bt,\vc x)^T$.
In (\ref{time-evol-mu}), time is measured in units of single particle updates.
It is clear that for sufficiently slow annealing and possibly after an initial burn-in period, the system will be approximately described by the equilibrium distribution (\ref{pi}), with $T^e$ replaced by $T^e(t)$.
If the annealing is too fast, however, convergence might again become slow, or the system might converge to a biased result.

It is intuitively clear that slow or even biased convergence can occur if the relaxational processes within the system (controlled by the jump distribution $k(\bt,\bt')$) are slow compared to the annealing speed.
For sufficiently fast mixing, on the other hand, we work under the {\em endoreversibility assumption}, which states that, at any time, the system is approximately described by an equilibrium distribution (\ref{pi}),
\begin{equation}\label{endoreversibility}
	\mu(\vc z,t)\approx \pi_{T(t)}(\vc z)\,,
\end{equation}
but with a {\em system temperature} $T(t)$ that is higher than the environmental temperature $T^e(t)$ used for the particle updates.
We will take this assumption for granted, for the time being, and confirm its validity at the end of this subsection.

To find an efficient annealing scheme, we need an information-theoretic framework.
Following \cite{andresen_1994_CTS},  particle updates are associated with a flow of information from the environment into the system, or, equivalently, as an entropy flow from the system to the environment.
Part of this entropy stems from the reduction of the system entropy when it transforms from the prior to the posterior. This part is path-independent and constitutes the well invested part of the computation.
The rest is the inevitable {\em entropy production}, which is computational waste we intend to minimize.
According to \cite{seifert_2005_EntropyProductionMasterEq}, entropy production can be understood as loss of information due to the rejections encountered during annealing relative to the inverse (heating) process.
It is thus a measure for the irreversibility of the computation.
However, there is also a "reversible computational waste": Running the algorithm at equilibrium does not produce any entropy. Nevertheless it is to be avoided as it does not lead to a reduction of the system entropy either.
This is due to the fact that we disregard the information stored in the history of each particle and only consider the current state.
Minimizing entropy production under the endoreversibility assumption will leave us with a family of algorithms that is essentially parameterized by two parameters governing, respectively, the mixing in parameter space and the annealing speed.
These parameters will have to be tuned in such a way that (i) the endoreversibility assumption is justified and (ii) there is not too much reversible computational waste.

Under the endoreversibility assumption (\ref{endoreversibility}) the entropy production rate, $\dot\sigma(t)$, is given by \cite{Albert_2013_ABC}
\begin{equation}\label{entprod}
 \dot\sigma(t) = \dot U(t)F(t)\,,
\end{equation}
where $U(t)$ denotes the system's energy\footnote{In \cite{Albert_2013_ABC}, the average particle distance was used. Here, we use the total distance because it is extensive in $N$.}
\begin{equation}\label{U}
U(t)=N\int u(\vc x)\mu(\vc z,t)d\vc z
\approx
\sum_{i=1}^N u(\vc x_i)
\,,
\end{equation}
and $F(t)$ the associated thermodynamic force
\begin{equation}
F(t)=\frac{1}{T(t)} - \frac{1}{T^e(t)}\,.
\end{equation}
Under fixed initial and final energies, an easy exercise in variational calculus \cite{spirkl_1995_optFiniteTimeEndorevProc} reveals the necessary and sufficient condition for minimal total entropy production to be given by eq.
\begin{equation}\label{Spirkl}
\frac{\partial F}{\partial \dot U}\dot U^2=v=\const\,.
\end{equation}
In order to derive an adaptive annealing schedule from (\ref{Spirkl}) we need to know both the functional dependence between $\dot U$ and $F$ and the functional dependence between $U$ and $T$.

In standard SA algorithms that are used as global optimizers, all the information is in the metric.
In our case, the metric is just an auxiliary construct and all the information lies in the likelihood function.
We are thus allowed to redefine the metric in order to design a more efficient algorithm.
If $\rho(\vc x,\vc y)$ is some user-defined metric (such as the one defined by the right eq. of (\ref{alpha})), we replace the left eq. of (\ref{alpha}) by eq.
\begin{equation}\label{newmetric}
u(\vc x):=\int_{\rho(\vc x',\vc y) \leq \rho(\vc x,\vc y)}
L(\vc x'|\bt)f_{pri}(\bt)d\bt d\vc x'\,.
\end{equation}
This new energy has a uniform prior density on the interval $[0,1]$, $f_{pri}(u)=\chi([0,1])$.
Thus, through a redefinition of the user-defined metric $\rho$ to the one defined in (\ref{newmetric}), we achieve a {\em specific heat capacity} that is approximately unity.
Indeed, w.r.t. this new metric, we see that, neglecting boundary terms of order $e^{-1/T(t)}$,
\begin{equation}\label{UT}
U(t)\approx NT(t)\,.
\end{equation}
The relationship between this new metric and the user-defined metric $\rho$ can be established, approximately, during the initialization of the algorithm, when the initial ensemble is generated \cite{Albert_2013_ABC}.

Using the new metric (\ref{newmetric}) we also find that, up to quadratic order in $T$ and $T^e$ \cite{Albert_2013_ABC},
\begin{equation}\label{Udot}
  \dot U(t) \approx \dot U(T,T^e)
  \approx
  -\gamma(T^2-(T^e)^2)\,,
\end{equation}
with
\begin{equation}\label{Udot2}
  \gamma
  =
  \left(
    \int L(\vc y|\bt)f_{pri}(\bt)d\bt
  \right)^{-2}
  \int k(\bt,\bt')L(\vc y|\bt)f_{pri}(\bt)L(\vc y|\bt')f_{pri}(\bt')d\bt d\bt'
 \,.
\end{equation}
Using (\ref{UT}) and (\ref{Udot}), criterion (\ref{Spirkl}) now reads as
\begin{equation}\label{Schedule}
\frac{(U(t)/N-T^e(t)^2)^2}{2T^e(t)^3}
=
\frac{v}{\gamma}\,.
\end{equation}
From (\ref{UT}), (\ref{Udot}) and (\ref{Schedule}) we derive the asymptotics of the annealing as \cite{Albert_2013_ABC}
\begin{equation}
T^e(t)\sim t^{-4/3}\,.
\end{equation}
We employ (\ref{Schedule}) adaptively, that is, we approximate $U(t)/N$ with the ensemble mean of the distances to the target, measured w.r.t. the new metric (\ref{newmetric}) and determine $T^e(t)$ solving the quartic equation (\ref{Schedule}).
The parameter $\gamma$ can either be determined, approximately, at the beginning of the algorithm, or we may treat $v/\gamma$ as our tuning parameter that governs the annealing speed.

The second tuning parameter of the algorithm governs the {\em mixing speed}.
We may simply choose, for the jump distribution $k(\bt,\bt')$, a normal distribution whose covariance matrix is given by
\begin{equation}
\label{beta2}
  \beta\Sigma(t)+s\tr(\Sigma(t))\identy\,,
\end{equation}
where $\beta$ is our tuning parameter and $\Sigma(t)$ an empirical covariance matrix that may be estimated from the $\bt$ components of the ensemble $E$ (either at the beginning only or throughout run-time, hence the time dependence).
The empirical covariance of the ensemble might degenerate. Therefore, we add the second term in (\ref{beta2}), where $s$ is some small constant, to make sure that the whole parameter space is explored.
Notice that our previous derivation assumed a time-independent jump distribution.
In certain cases, however, it might be beneficial to adapt it to the empirical covariance of the ensemble throughout run-time \cite{Albert_2013_ABC}.

Before going to the general case we show that, in the limit of infinitely fast mixing and infinitely fast annealing assumption (\ref{endoreversibility}) is consistent with eqs. (\ref{transition}) and (\ref{time-evol-mu}).
Infinitely fast mixing means that instead of jumping in parameter space we simply draw from the prior, and, since the prior is assumed to be flat, this is achieved by setting $k(\bt,\bt')=1$.
In this case only the marginal density of the energy $u$ is affected by particle updates.
Under the endoreversibility assumption (\ref{endoreversibility}), and w.r.t. the new metric (\ref{newmetric}), this density reads
\begin{equation}\label{endoreversibility1}
  \mu(u;t)
  =
  \frac{1}{T(t)}
  e^{-u/T(t)}\,.
\end{equation}
Plugging (\ref{endoreversibility1}) into the master equation (\ref{time-evol-mu}) yields, up to terms of order $\exp[-1/T(t)]$,

\begin{multline}\label{mudot}
  \dot\mu(t;u)
  =
  \frac{1}{T(t)}
  \int
  \bigg(
    \min\left(
        1,\exp\left[-\frac{u-u'}{T^e(t)}\right]
    \right)
    \exp\left[-\frac{u'}{T(t)}\right]
    -\\
    \min\left(
        1,\exp\left[-\frac{u'-u}{T^e(t)}\right]
    \right)
    \exp\left[-\frac{u}{T(t)}\right]
  \bigg)
  du'
  \\
  =
  \frac{e^{-u/T(t)}}{T(t)}
  \bigg\lbrace
    \int_0^u
    \left(
        \exp\left[
            -(u'-u)\left(\frac{1}{T(t)}-\frac{1}{T^e(t)}\right)
        \right]
        -1
    \right)
    du'\\
    +
    \int_u^\infty
    \left(
        \exp\left[
            -\frac{(u'-u)}{T(t)}
        \right]
        -
        \exp\left[
            -\frac{(u'-u)}{T^e(t)}
        \right]
    \right)
    du'
    \bigg\rbrace
    \\
    =
    \frac{e^{-u/T(t)}}{T(t)}
  \bigg\lbrace
    \int_0^u
        \exp\left[
            -u'\left(\frac{1}{T^e(t)}-\frac{1}{T(t)}\right)
        \right]du'
    -u\\
    +
    \int_0^\infty
    \left(
        \exp\left[
            -\frac{u'}{T(t)}
        \right]
        -
        \exp\left[
            -\frac{u'}{T^e(t)}
        \right]
    \right)
    du'
    \bigg\rbrace
    \\
    =
    \frac{e^{-u/T(t)}}{T(t)}
    \bigg\lbrace
        T(t)-T^e(t)-u
        +
        \frac{T(t)T^e(t)}{T(t)-T^e(t)}
        \left(
            1-\exp\left[
                -u\frac{T(t)-T^e(t)}{T(t)T^e(t)}
            \right]
        \right)
    \bigg\rbrace\,.
\end{multline}
On the other hand, the temporal derivative of (\ref{endoreversibility1}) reads
\begin{equation}\label{mudot1}
  \dot\mu(t;u)
  =
  -\frac{e^{-u/T(t)}\dot T(t)}{T^3(t)}(T(t)-u)\,.
\end{equation}
Eqs. (\ref{mudot}) and (\ref{mudot1}) are consistent in the limit of infinitely fast annealing, $T^e(t)\equiv 0$, upon setting $\dot T(t)=-T^2(t)$, i.e., for $T(t)\sim 1/t$.
This calculation also shows that, if annealing has a finite speed, there is additional internal entropy production due to a violation of the endoreversibility assumption.
However, unless annealing is extremely slow, we expect (\ref{entprod}) to be the dominant source of entropy production.
Notice that, as long as mixing in parameter space is sufficiently fast and the final temperature is sufficiently low, the $\bt$-distribution represented by our final ensemble will be a faithful representation of the posterior.
A violation of the endoreversibility assumption only means that we might not have found the most efficient annealing schedule.

\subsection{The general case}

If the prior carries relevant information, we must make sure that it is appropriately represented in the posterior.
If we expect to learn little, i.e., if we expect the posterior not to differ too much from the prior, we can simply employ a rejection-ABC \cite{tavare_1997_ABC, weiss_1998_ABC}.
That is, we iteratively simulate from the joint prior $f_{pri}(\bt)L(\vc x|\bt)$ and accept only those $\bt$ whose corresponding $\vc x$ are close enough to the data $\vc y$ (w.r.t. a fixed and small tolerance).
This algorithm becomes inefficient if the posterior differs a lot form the prior, in which case we better employ an algorithm that is based on random jumps in parameter space.
In that case, we must make sure that prior and likelihood decide on equal footing whether a move is accepted or not.

We address this problem introducing a second energy variable,
\begin{equation}
  u_2(\bt)=-\ln\left(f_{pri}(\bt)\right)\,,
\end{equation}
and replacing transition rate (\ref{transition}) by \footnote{We use a notation that is somewhat different from \cite{Albert_2013_ABC}, in order to comply with physics' notation.}
\begin{equation}\label{newtransition}
q_{{\vc T}^e}((\bt',\vc x')\rightarrow(\bt,\vc x))=
k(\bt',\bt)L(\vc x|\bt)
\min
\bigg(1,
	\exp\left[
        -\frac{u_1(\vc x)-u_1(\vc x')}{T_1^e}
        -\frac{u_2(\bt)-u_2(\bt')}{T_2^e}
    \right]
\bigg)\,,
\end{equation}
where $u_1(\vc x)$ denotes some user-defined distance to the measured data, such as the one given in eq. (\ref{alpha}).
Transition (\ref{newtransition}) satisfies a detailed balance condition analogous to (\ref{detbal}), but for the equilibrium distribution
\begin{equation}\label{newpi}
  \pi_{{\vc T}^e}(\bt,\vc x)
  =
  Z^{-1}({\vc T}^e)
  L(\vc x|\bt)
  e^{-u_1(\vc x)/T_1^e-u_2(\bt)/T_2^e}\,,
\end{equation}
with
\begin{equation}
  Z({\vc T}^e)
  =
  \int
  L(\vc x|\bt)
  e^{-u_1(\vc x)/T_1^e-u_2(\bt)/T_2^e}
  d\bt d\vc x\,.
\end{equation}
Analogously to Sect. \ref{SectUninformative}, for sufficiently fast mixing, we make the endoreversibility assumption,
\begin{equation}\label{endoreversibility2}
  \mu(\vc z,t)\approx
  \pi_{\vc T}(\vc z,t)\,,
\end{equation}
and expect the dominant source of entropy production to be given by eq.
\begin{equation}\label{heatflow1}
\dot\sigma(t) = \vc F(t)^T\dot{\vc U}(t)\,,
\end{equation}
with the extensities
\begin{align}
  U_1(t)&:=N\int u_1(\vc x)\mu(\bt, \vc x,t)d\bt d\vc x
  \approx\sum_{i=1}^N u_1(\vc x_i)
  \,,\label{U1}\\
  U_2(t)&:=N\int u_2 (\bt  )\mu(\bt, \vc x,t)d\bt d\vc x
  \approx\sum_{i=1}^N u_2(\bt_i)
  \,,\label{U2}
\end{align}
and the thermodynamic forces
\begin{equation}
\vc F(t)
=
\begin{pmatrix}
T_1(t)^{-1} - T^e_1(t)^ {-1}\\
T_2(t)^{-1} - T^e_2(t)^ {-1}\,
\end{pmatrix}\,.
\end{equation}
The two-dimensional problem does not lend itself to a simple redefinition of the metric, as in the previous section.
Therefore, we can only uphold the endoreversibility assumption (\ref{endoreversibility2}), for sufficiently small thermodynamic forces.

A necessary criterion for minimal entropy production, for fixed initial and final values of $\vc U$, is given by \cite{spirkl_1995_optFiniteTimeEndorevProc}
\begin{equation}\label{preschedule2}
 \dot{\vc U} ^T \frac{\partial \vc F}{\partial \dot{\vc U}}
 \dot{\vc U} = v = \const\,.
\end{equation}
To derive an annealing schedule from (\ref{preschedule2}) we need to know the functional dependencies between $\vc U$ and $\vc T$ and between $\dot{\vc U}$ and $\vc F$, respectively.
For sufficiently small thermodynamic forces, we make the linearity assumption
\begin{equation}\label{Onsager}
\dot {\vc U}(t) = L(\vc U(t))\vc F(t)\,,
\end{equation}
where $L$ denotes the {\em Onsager} matrix (which is {\em symmetric} due to detailed balance).
Under assumption (\ref{Onsager}), condition (\ref{preschedule2}) leads to a constant entropy production rate \cite{spirkl_1995_optFiniteTimeEndorevProc}
\begin{equation}\label{constentropyprod}
  \dot\sigma(t)=
  \dot{\vc U} ^T R(\vc U)
 \dot{\vc U} = v\,,
\end{equation}
where $R(\vc U):=L^{-1}(\vc U)$ defines a metric on the $(U_1,U_2)$-plane.
Due to the Cauchy-Schwartz inequality
\begin{equation}\label{sigma}
  \sigma =\int_{t_i}^{t_f}
  \dot{\vc U}(t) ^T R(\vc U(t)) \dot{\vc U}(t) dt
  \geq
  \frac{\mathcal K}{t_f - t_i}\,,
\end{equation}
where $\mathcal K$ is the length of the process-path in the $(U_1,U_2)$-plane, measured with the metric $R(\vc U)$.
The lower bound of (\ref{sigma}) is assumed if the integrand is constant,
i.e., if the entropy production rate is constant.
Thus, under linearity assumption (\ref{Onsager}), a necessary and sufficient condition for minimal entropy production is to travel along a path of minimal length (w.r.t. metric $R$) in such a way that the entropy production rate is constant \cite{salamon_1983_ThdynLengthDissipAvailability}.

To derive an adaptive annealing schedule from this we continuously need to estimate the Onsager matrix $L(\vc U)$, during run-time.
To this end, we derive from (\ref{time-evol-mu}), (\ref{newtransition}), (\ref{endoreversibility2}), (\ref{U1}) and (\ref{U2}) the {\em fluctuation-dissipation theorem}
\begin{multline}\label{L}
  L_{ij}(\vc U)=Z^{-1}({\vc T})
  \int
  (u_i(\vc z)-u_i(\vc z'))
  (u_j(\vc z)-u_j(\vc z'))
  \\ \times
  k(\bt,\bt')
  L(\vc x|\bt)L(\vc x'|\bt')
  \exp[
    -u_1(\vc x)/T_1-u_2(\bt)/T_2
    ]
   \\ \times
  \chi\left(
    (u_1(\vc x)-u_1(\vc x'))/T_1+(u_2(\bt)-u_2(\bt'))/T_2
    \right )
    \\ \times
    d\vc xd\vc x'd\bt d\bt'\,.
\end{multline}
The r.h.s. of (\ref{L}) can be estimated, during run-time, in different ways: It can be estimated, in a rather straightforward way, using both the ensemble $E$ and the prior sample that is generated, as a side product, when the initial ensemble is generated \cite{Albert_2013_ABC}. In a later stage of the algorithm, we can also estimate it by means of a matrix of attempted moves that has been populated during run-time \cite{Albert_2013_ABC, andresen_1988_lumpedThdynPropSA}.

Since, at the beginning of the algorithm, neither is the target value for $U_2$, at $T_1=0$ and $T_2=1$, known exactly nor is the metric $R(\vc U)=L^{-1}(\vc U)$ known globally. Thus, it appears difficult to come up with an optimal path in the $(U_1,U_2)$-plane up-front.
Therefore, we force the process to be on a path such that $T_2$ remains close to unity, at all times.
Then, the prior always has the correct weight.
Practically, this can be achieved by applying a counter force, setting
\begin{equation}
  \frac{1}{T_2^e}-1
  =
  -a\left(
        \frac{1}{T_2}-1
    \right)
    \,,
\end{equation}
where $a$ is some positive tuning parameter.

The last ingredient for the algorithm is a run-time estimate of the system's intensities $\vc T$, based on the measured extensities $\vc U$.
To this end we derive from (\ref{endoreversibility2}), (\ref{U1}) and (\ref{U2}) the {\em fluctuation-dissipation theorem}
\begin{equation}\label{Jacobi}
      \frac{\partial \vc U}{\partial{\vc T}}
      =
      N
      \begin{pmatrix}
       \frac{1}{T_1^2}\Var(u_1)
            &   \frac{1}{T_2^2}\Cov(u_1,u_2) \\
       \frac{1}{T_1^2}\Cov(u_1,u_2)
            &   \frac{1}{T_2^2}\Var(u_2)
     \end{pmatrix}\,.
\end{equation}
The r.h.s. of (\ref{Jacobi}) can easily be estimated with the empirical covariance of the ensemble $E$ and used to estimate the change in intensities, $\Delta{\vc T}$, following a measured change of extensities, $\Delta\vc U$, by means of the approximation
\begin{equation}
\label{Delta-eps}
\Delta{\vc T}
\approx
\left(
    \frac{\partial\vc U}{\partial{\vc T}}
\right)^{-1}
\Delta\vc U\,.
\end{equation}
Of course, occasionally, the estimate of $\vc T$ needs to be corrected using the endoreversibility assumption (\ref{endoreversibility2}) \cite{Albert_2013_ABC}.

\section{Conclusions}

We have presented a scheme of algorithms that allow us to extract probabilistic information about model parameters from data.
The scheme is generic in the sense that we only need access to the model in the form of a {\em simulator}, i.e., we do not need access to the {\em density} of the model's likelihood function.

Our scheme is inspired by a body of literature on adaptive simulated annealing \cite{nulton1988statistical, salamon1988simulated, andresen_1988_lumpedThdynPropSA, ruppeiner_1991_EnsembleSA, andresen_1994_CTS}.
To design an efficient annealing schedule we employ the principle of minimal entropy production \cite{salamon1988simulated, nulton1988statistical}.
Our algorithms are adaptive as they determine the state of the system from measured thermodynamic extensive quantities.
The underlying {\em endoreversibility assumption} (i.e. the assumption that the state of the system is approximately described by a Gibbs state w.r.t. the measured extensities) is approximately satisfied as long as the mixing within the system (determined by the jump width) is sufficiently fast and the annealing speed sufficiently slow.
Under the endoreversibility assumption, entropy production is given by the flow of extensities from the system to the environment (or vice versa). Minimization of this entropy production leads to a family of adaptive annealing schedules that are basically parameterized by two parameters, determining, respectively, the mixing and the annealing speed.

In contrast to optimization the complexity of our task is in the likelihood function, from which we only have to draw model outputs, and not in the metric that is used to accept or reject a suggested move in parameter space.
In the practically relevant case of negligible prior information, a simple redefinition of the metric allows us to go beyond the linearity assumption that underlies many optimization algorithms and derive an optimal algorithm for fast annealing, which is not based on a constant entropy production rate.
Furthermore, in this special case, the bias in the final result will be small as long as mixing in parameter space is sufficiently fast and the final temperature is sufficiently low.

In the general case of non-negligible prior knowledge we need to take care that the prior is properly represented in the posterior. We do this by introducing a second energy, measuring prior energy of a particle, and an associated temperature $T_2^e$ (control parameter).
For that general case, we can only derive an optimal annealing schedule that is sufficiently slow to justify both the endoreversibility assumption as well as a linearity assumption between thermodynamic forces and fluxes. Under this assumption, optimality leads to a constant entropy production rate \cite{spirkl_1995_optFiniteTimeEndorevProc}.
While the adaptation of $T_2^e$ aims at a proper representation of the prior, a violation of the endoreversibility assumption might lead to a bias nonetheless.
If the prior is very strong, in order to avoid a bias, annealing might have to be so slow that the algorithm becomes inferior to a simple rejection ABC algorithm \cite{tavare_1997_ABC, weiss_1998_ABC}.
This deficiency is inherent to all ABC algorithms that are based on a decreasing tolerance.

As all ABC algorithms are based on drawing outputs from the likelihood function their efficiency drops drastically with the dimension of the output space.
If the dimension of the output space is large, but the dimension of the parameter space small it is often possible to reduce the output dimension to the order of the dimension of the parameter space without loosing too much information about the parameters. A semi-automatic method to produce these so-called {\em summary statistics} has been proposed in \cite{Fearnhead_2012_ABC}.

An implementation in the free statistics software R\footnote{http://cran.r-project.org/} of the methods described in this paper, including semi-automatic summary statistics, is available for free download from the author's github\footnote{https://github.com/carloalbert/sabc/tree/master/SABC} and as part of the R-package Easy-ABC\footnote{http://cran.r-project.org/web/packages/EasyABC/index.html}.
Furthermore, pseudocodes of the suggested algorithms along with several case studies can be found in the original publication \cite{Albert_2013_ABC}.

\section*{Acknowledgement}
  The author wishes to acknowledge the inspiring atmosphere during the 2015 workshop on energy landscapes at the Telluride Science Research Center (TSRC).

\bibliographystyle{plain}
\bibliography{C:/Users/albertca/SWITCHdrive/refs}

\end{document}